\documentclass[letterpaper,preprintnumbers,prd,nofootinbib,nobibnotes,showpacs]{revtex4}
\usepackage{amsfonts}
\usepackage{mathrsfs}
\usepackage{epsfig}
\usepackage{graphicx}%
\usepackage{dcolumn}
\usepackage{amsmath}
\usepackage{color}
\usepackage{color}
\makeatletter
\def\btt#1{\texttt{\@backslashchar#1}}%
\DeclareRobustCommand\bblash{\btt{\@backslashchar}}%
\makeatother
\begin{document}

\title{Extending Horndeski theories into Lovelock gravity}
\author{Changjun Gao}\email{gaocj@bao.ac.cn} \affiliation{ Key Laboratory of Computational Astrophysics, National Astronomical Observatories, Chinese
Academy of Sciences, Beijing 100012, China}
\affiliation{University of Chinese Academy of Sciences, Beijing 100049, China}

\date{\today}

\begin{abstract}
 The Horndeski theories are extended into the Lovelock gravity theory.  When the canonical scalar field is uniquely kinetically coupled to the Lovelock tensors, it is named after Lovelock scalar field. The Lovelock scalar field model is a subclass of the new Horndeski theories.  A most attractive feature of the Lovelock
 scalar field is its equation of motion is second order. So it is free of ghosts. We study the cosmology of Lovelock scalar field in the background of $7$ dimensional spacetime and present a class of cosmic solutions. These solutions reveal the physics of the scalar field is rather rich and merit further study.
\end{abstract}



\maketitle

\section{Introduction}

The Horndeski theories are the most general scalar-tensor theories with second-order equations of
motion \cite{horn:1974} which guarantees it is free of ghosts. The theories have not attracted much attention for a long time
until the discoveries of covariant Galileons \cite{nic:2008,deff:2009} and
generalized Galileon theories \cite{charm:2012, deff:2011}.
It is found the Horndeski theories contain a wide variety of gravitational theories such as, General Relativity with a minimally coupled scalar field functions
\cite{arm:1999}, the Brans-Dicke theory \cite{bran:1961}, the dilaton gravity theory \cite{gas:1993}, the covariant Galileons \cite{deff:2009}, the derivative coupling \cite{am:1993, ger:2010} and the Gauss-Bonnet coupling
\cite{kob:2011}. Furthermore, the Horndeski theories have been extended into multi-field case \cite{ext:1,ext:2,ext:3,ext:4,ext:5}. The extending of sing-field Horndeski theory can be found in Refs.~\cite{exts:1,exts:2,exts:3,exts:4,exts:5,exts:6,exts:7,exts:8}.

The purpose of this paper is to extend the sing-field Horndeski theories into the Lovelock gravity. The Lovelock gravity is the most general gravity theory with second-order equations of motion \cite{ll:1971}. Same as the Einstein tensor $G_{\mu\nu}$, the Lovelock tensors $G_{\mu\nu}^{(p)}$ are also second order and obey the divergence-free law of $G^{(p);\nu}_{\mu\nu}=0$. It is this desired property that enables us to generalize the Horndeski theories. The paper is organized as follows. In section~\ref{sec:2}, we make a brief review of Lovelock theory in advance and then extend the Horndeski theories within it. In
section~\ref{sec:3}, we investigate the cosmology of Lovelock scalar field in the background of $7$ dimensional spacetime. In $4$ dimensional spacetime, the Lovelock tensors higher than first order (Einstein tensor) are all vanishing. So in order to observe the effect of $G_{\mu\nu}^{(2)}$ and $G_{\mu\nu}^{(3)}$, we must work in at least $7$ dimensional spacetime. In section~\ref{sec:4}, a class of cosmic solutions are presented numerically. Finally, conclusion and discussion are given in section~\ref{sec:5}. Throughout this paper, we adopt the system of units in which $G=c=\hbar=1$ and the metric signature
$(-,\ +,\ +,\ +)$.
\section{Generalize the Horndeski theories}\label{sec:2}
The Horndeski theories gives the most general second-order equation of motion for scalar field in four dimensional spacetime. This theory is described by the Lagrangian \cite{deff:2011}
\begin{eqnarray}
L=\sum_{i}^{5}L_i\;,
\end{eqnarray}
where
\begin{eqnarray}
L_2&=&G_2\left(\phi,\ X\right)\;,\\
L_3&=&G_3\left(\phi,\ X\right)\square X\;,\\
L_4&=&G_4\left(\phi,\ X\right)R-2G_{4X}\left(\phi,\ X\right)\left[\left(\square\phi\right)^2-\phi^{;\mu\nu}\phi_{;\mu\nu}\right]\;,\\
L_5&=&G_5\left(\phi,\ X\right)G_{\mu\nu}\phi^{;\mu\nu}
+\frac{1}{3}G_{5X}\left(\phi,\ X\right)\left[\left(\square\phi\right)^3-3\square\phi\phi_{;\mu\nu}\phi^{;\mu\nu}+2\phi_{;\mu\nu}\phi^{;\mu\sigma}\phi^{;\nu}_{;\sigma}\right]\;.
\end{eqnarray}
Here $G_i$ are functions in terms of the scalar field $\phi$ and its kinetic
energy $X\equiv\partial_{\mu}\phi\partial^{\mu}\phi/2$. $\phi^{;\mu\nu}\equiv \nabla^{\mu}\nabla^{\nu}\phi$, $G_{iX}\equiv \partial G_i/\partial X$ and
$G_{i\phi}\equiv\partial G_i/\partial\phi$. $R$ is the Ricci scalar and $G_{\mu\nu}$ is the Einstein tensor.
Horndeski derived the Lagrangian of the most general scalar-tensor theories
in a different form \cite{horn:1974} in 1974. But Ref.\cite{kob:2011} shows that it is equivalent to the above
form.

In order to extend the Horndeski theories into the Lovelock gravity, let' s make a brief review on the Lovelock gravity theory. Starting from the vacuum Einstein-Hilbert action
\begin{eqnarray}
{S}=\int d^4 x\sqrt{-g}R=\int d^4 x\sqrt{-g}\delta_{\sigma_1\sigma_2}^{\lambda_1\lambda_2}
R_{\lambda_1\lambda_2}^{\sigma_1\sigma_2}\;,
\end{eqnarray}
one obtain the Einstein equations for gravity
\begin{eqnarray}
G_{\mu\nu}=0\;.
\end{eqnarray}
Here $\delta_{\sigma_1\sigma_2}^{\lambda_1\lambda_2}$ is the generalized Kronecker delta function of order two, $R_{\lambda_1\lambda_2}^{\sigma_1\sigma_2}$ is the Riemann tensor. The Einstein equations
are second-order differential equations and $G_{\mu\nu}$ has the property of vanishing divergence, $G_{\mu;\nu}^{\nu}=0$.

A remarkable property of gravitational invariant is its linearity with respect to the Riemann tensor.  Lovelock \cite{ll:1971} suggested to give up this requirement
and extend the invariants to power laws of Riemann tensor
\begin{eqnarray}
L_{p}=2^{-p}\delta_{\sigma_1\sigma_2\cdot\cdot\cdot \sigma_{2p}}^{\lambda_1\lambda_2\cdot\cdot\cdot \lambda_{2p}}
R_{\lambda_1\lambda_2}^{\sigma_1\sigma_2}R_{\lambda_3\lambda_4}^{\sigma_3\sigma_4}\cdot\cdot\cdot R_{\lambda_{2p-1}\lambda_{2p}}^{\sigma_{2p-1}\sigma_{2p}}
\;,
\end{eqnarray}
where $\delta_{\sigma_1\sigma_2\cdot\cdot\cdot \sigma_{2p}}^{\lambda_1\lambda_2\cdot\cdot\cdot \lambda_{2p}}$
is the generalized Kronecker delta of the order $2p$. It equals to $\pm 1$ if the upper indices form
an even or odd permutation of the lower ones, respectively,
and zero in all other cases. Then the Lovelock action of gravity takes the form
\begin{eqnarray}
{S}=\int d^n x\sqrt{-g}\sum_{p}\alpha_p L_{p}\;,\label{ll}
\end{eqnarray}
where $n$ is the dimension of spacetime, $\alpha_p$ are constants and
summation is carried over all $p$.

From Eq.~(\ref{ll}) one obtain the Lovelock gravitational equations
\begin{eqnarray}
\sum_{p}\alpha_p G^{(p)}_{\mu\nu}=0\;,
\end{eqnarray}
with
\begin{eqnarray}
G^{(p)}_{\mu\nu}=g_{\mu\rho}\delta_{\nu\sigma_1\sigma_2\cdot\cdot\cdot \sigma_{2p}}^{\rho\lambda_1\lambda_2\cdot\cdot\cdot \lambda_{2p}}
R_{\lambda_1\lambda_2}^{\sigma_1\sigma_2}R_{\lambda_3\lambda_4}^{\sigma_3\sigma_4}\cdot\cdot\cdot R_{\lambda_{2p-1}\lambda_{2p}}^{\sigma_{2p-1}\sigma_{2p}}\;,
\end{eqnarray}
such that \cite{bri:1997}
\begin{eqnarray}
&&G^{(0)}_{\mu\nu}=g_{\mu\nu}\;,\nonumber\\
&&G^{(1)}_{\mu\nu}=G_{\mu\nu}\;,\nonumber\\
&&G^{(2)}_{\mu\nu}=-\frac{1}{2}g_{\mu\nu}\left(R^2-4R_{\kappa\sigma}R^{\kappa\sigma}+R_{\kappa\sigma\tau\rho}
R^{\kappa\sigma\tau\rho}\right)+2\left(RR_{\mu\nu}-R_{\mu\sigma\kappa\tau}R^{\kappa\tau\sigma}_{\nu}-2R_{\mu\kappa\nu\sigma}R^{\kappa\sigma}
-2R_{\mu\sigma}R^{\sigma}_{\nu}\right)\;,\nonumber\\
&&G^{(3)}_{\mu\nu}=\frac{1}{2}g_{\mu\nu}\left(12RR_{\kappa\sigma}R^{\kappa\sigma}-R^3-3RR_{\alpha\beta\sigma\kappa}R^{\alpha\beta\sigma\kappa}
-16R_{\alpha}^{\beta}R_{\beta}^{\sigma}R_{\sigma}^{\alpha}+24R_{\alpha\beta}R_{\sigma\kappa}R^{\alpha\sigma\beta\kappa}+24R_{\alpha}^{\beta}
R^{\alpha\sigma\kappa\rho}R_{\beta\sigma\kappa\rho}\right.\nonumber\\&&\left.+2R_{\alpha\beta}^{\sigma\kappa}R_{\sigma\kappa}^{\rho\lambda}R_{\rho\lambda}^{\alpha\beta}
-8R_{\alpha\beta}^{\sigma\kappa}R_{\sigma\rho}^{\alpha\lambda}R_{\kappa\lambda}^{\beta\rho}\right)-24R_{\mu\alpha\beta\sigma}
R_{\nu}^{\beta}R^{\alpha\sigma}
-12R_{\mu\nu}R_{\alpha\beta}R^{\alpha\beta}+24R_{\mu}^{\alpha}R_{\alpha}^{\beta}R_{\beta\nu}
+24R_{\mu}^{\alpha}R^{\beta\sigma}R_{\alpha\beta\sigma\nu}\nonumber\\&&+3R_{\mu\nu}R^2+3R_{\mu\nu}R_{\alpha\beta\sigma\kappa}R^{\alpha\beta\sigma\kappa}
-12R_{\mu\alpha}R_{\nu\beta\sigma\kappa}R^{\alpha\beta\sigma\kappa}
+6RR_{\mu\alpha\beta\sigma}R_{\nu}^{\alpha\beta\sigma}-24R_{\mu\alpha\nu\beta}R_{\sigma}^{\alpha}R^{\sigma\beta}-12RR_{\mu}^{\sigma}R_{\sigma\nu}\nonumber\\&&+24R_{\mu\alpha\nu\beta}R_{\sigma\kappa}R^{\alpha\sigma\beta\kappa}-12R_{\mu\alpha\beta\sigma}
R^{\kappa\alpha\beta\sigma}R_{\kappa\nu}-12R_{\mu\alpha\beta\sigma}R^{\alpha\kappa}R_{\nu\kappa}^{\beta\sigma}
+12RR_{\mu\sigma\nu\kappa}R^{\sigma\kappa}\nonumber\\&&+12R_{\mu\alpha\nu\beta}R^{\alpha}_{\sigma\kappa\rho}R^{\beta\sigma\kappa\rho}+
6R_{\mu}^{\alpha\beta\sigma}R_{\beta\sigma}^{\kappa\rho}R_{\kappa\rho\alpha\sigma}+24R_{\mu\alpha}^{\beta\sigma}R_{\beta\nu\rho\lambda}
R_{\sigma}^{\lambda\alpha\rho}+24R_{\mu}^{\alpha\beta\sigma}R_{\beta}^{\kappa}R_{\sigma\kappa\nu\beta}\;,\nonumber\\
&&G^{(4)}_{\mu\nu}=\cdot\cdot\cdot\cdot\cdot\cdot\;.
\end{eqnarray}
The order $p$ of the Lovelock tensor is related to the dimension $n$ of spacetime by $2p\leq n-1$. Same as the Einstein equations, the Lovelock equations are also second order which guarantees the absence of ghosts \cite{zz:1985}. On the contrary,
fourth order modified gravities are usually plagued by ghost problem \cite{sf:2010}. Since the Lovelock tensors $G^{(p)}_{\mu\nu}$ are derived by variations of the action, they exactly obey the law of $G^{(p);\nu}_{\mu\nu}=0$. The desired property of up to second order derivative and free of divergence for Lovelock tensors leads us to generalize the Horndeski theory as follows

\begin{eqnarray}
L=\sum_{i}^{7}L_i\;,
\end{eqnarray}
where
\begin{eqnarray}
L_2&=&G_2\left(\phi,\ X\right)\;,\\
L_3&=&G_3\left(\phi,\ X\right)\square X\;,\\
L_4&=&G_4\left(\phi,\ X\right)R-2G_{4X}\left(\phi,\ X\right)\left[\left(\square\phi\right)^2-\phi^{;\mu\nu}\phi_{;\mu\nu}\right]\;,\\
L_5&=&G_5\left(\phi,\ X\right)G_{\mu\nu}\phi^{;\mu\nu}
+\frac{1}{3}G_{5X}\left(\phi,\ X\right)\left[\left(\square\phi\right)^3-3\square\phi\phi_{;\mu\nu}\phi^{;\mu\nu}+2\phi_{;\mu\nu}\phi^{;\mu\sigma}\phi^{;\nu}_{;\sigma}\right]\;,\\
L_6&=&G_6\left(\phi\right)\sum_{p}\alpha_p L_{p}\;,\\
L_7&=&G_7\left(\phi\right)\sum_{p}\beta_p G^{(p)}_{\mu\nu}\phi^{;\mu}\phi^{;\nu}\;.\\
\end{eqnarray}
Here $\beta_p$ are coupling constants.  The term $L_6$ can play an important role in the evolution of universe. For example, Ref.~\cite{noj:2005} investigated the situation of $L_6=G_6\left(\phi\right)L_{2}$ where it is named after Gauss-Bonnet dark energy.  On the other hand, the term $L_7$ can demonstrate its important role in higher dimensional physics. We shall see this point in the next section. In all, $L_6$ and $L_7$ are not trivial. The resulting equation of motion from (13) is exactly seconde-order for arbitrary dimensional spacetime. In the next, for simplicity, we focus on the subclass of
\begin{eqnarray}
G_2=2V\left(\phi\right)\;,\ \ G_3=0\;,\ \ G_4=0\;,\ \ G_5=0\;,\ \ G_6=0\;,\ \ G_7=1\;.
\end{eqnarray}
Then the corresponding Lagrangian is
\begin{equation}
\mathscr{L}=\sum_{p}\beta_p G^{(p)}_{\mu\nu}\nabla^{\mu}\phi\nabla^{\nu}\phi+2V\left(\phi\right)\;,
\end{equation}
where $V(\phi)$ is the scalar potential. We name this scalar field as Lovelock scalar field. The equation of motion is
\begin{equation}
\sum_{p}\beta_p G^{(p)}_{\mu\nu}\nabla^{\mu}\nabla^{\nu}\phi+V_{,\phi}=0\;.
\end{equation}
It is a equation of second order derivative. In four dimensional spacetime, we have $G^{(p)}_{\mu\nu}=0$ with $p\geq 2$. Therefore, in order to observe the effect of  $G^{(2)}_{\mu\nu}$ and $G^{(3)}_{\mu\nu}$, we should work in at least, $7$ dimensional spacetime because of $2p\leq n-1$.

\section{cosmic evolution}
\label{sec:3}
In this section, we shall investigate the cosmic evolution of scalar field in $7$ dimensional spacetime. The metric is given by
 \begin{eqnarray}
ds^2&=&-dt^2+a^2\left(t\right)\left(dx_1^2+dx_2^2+dx_3^2\right)+b^2\left(t\right)\left(dx_4^2+dx_5^2+dx_6^2\right)\;,
\end{eqnarray}
where $a(t)$ and $b(t)$ are the scale factor of our space and the extra space. In $7$ dimensional spacetime, the Lovelock tensors higher than third order are all vanishing. Thus the total action is give by
\begin{eqnarray}
{S}=\int d^7 x\sqrt{-g}\sum_{p=0}^{3}\left[\alpha_p L_{p}+\beta_p G^{(p)}_{\mu\nu}\nabla^{\mu}\phi\nabla^{\nu}\phi+2V\left(\phi\right)\right]\;.\label{ll7}
\end{eqnarray}
Then the equations of motion are
\begin{eqnarray}\label{eom1}
&&-\frac{1}{2}\alpha_0+\alpha_1\left(2\dot{H}+3H^2+3\dot{h}+6h^2+6Hh\right)
+\alpha_2\left[\left(48Hh+12H^2+12h^2\right)\dot{h}+\left(24Hh+24h^2\right)\dot{H}\right.\nonumber\\&&\left.+24H^3h+12h^4+72H^2h^2+72Hh^3\right]
-\alpha_3\left[288Hh^3\dot{H}+432H^2h^2\dot{h}+432H^2h^4+288H^3h^3\right]\nonumber\\&&
=\frac{1}{4}\beta_0\dot{\phi}^2+\beta_1\left[\left(\dot{H}+\frac{3}{2}H^2+\frac{3}{2}\dot{h}+3h^2+3Hh\right)\dot{\phi}^2+\left(2H+3h\right)\ddot{\phi}\dot{\phi}\right]
+\beta_2\left[\left(18h^2\dot{h}+18h^4+72Hh\dot{h}
\right.\right.\nonumber\\&&\left.\left.+36h^2\dot{H}+108H^2h^2+18H^2\dot{h}+36Hh\dot{H}+36H^3h
+108Hh^3\right)\dot{\phi}^2+\left(72Hh^2+12h^3+36H^2h\right)\ddot{\phi}\dot{\phi}\right]\nonumber\\&&
-\beta_3\left[\left(1080H^2h^2\dot{h}+1080H^2h^4+720Hh^3\dot{H}
+720H^3h^3\right)\dot{\phi}^2+720H^2h^3\ddot{\phi}\dot{\phi}\right]+V\;,
\end{eqnarray}
\begin{eqnarray}\label{eom2}
&&-\frac{1}{2}\alpha_0+\alpha_1\left(2\dot{h}+3h^2+3\dot{H}+6H^2+6Hh\right)
+\alpha_2\left[\left(48Hh+12h^2+12H^2\right)\dot{H}+\left(24Hh+24H^2\right)\dot{h}\right.\nonumber\\&&\left.
+24h^3H+12H^4+72H^2h^2+72hH^3\right]
-\alpha_3\left[288hH^3\dot{h}+432H^2h^2\dot{H}+432h^2H^4+288h^3H^3\right]\nonumber\\&&
=\frac{1}{4}\beta_0\dot{\phi}^2+\beta_1\left[\left(\dot{h}+\frac{3}{2}h^2+\frac{3}{2}\dot{H}+3H^2+3Hh\right)\dot{\phi}^2
+\left(2h+3H\right)\ddot{\phi}\dot{\phi}\right]+\beta_2\left[\left(18H^2\dot{H}+18H^4+72Hh\dot{H}
\right.\right.\nonumber\\&&\left.\left.+36H^2\dot{h}+108H^2h^2+18h^2\dot{H}+36Hh\dot{h}+36h^3H
+108hH^3\right)\dot{\phi}^2+\left(72hH^2+12H^3+36h^2H\right)\ddot{\phi}\dot{\phi}\right]\nonumber\\&&
-\beta_3\left[\left(1080H^2h^2\dot{H}+1080h^2H^4+720hH^3\dot{h}
+720H^3h^3\right)\dot{\phi}^2+720h^2H^3\ddot{\phi}\dot{\phi}\right]+V\;,
\end{eqnarray}
\begin{eqnarray}\label{eom3}
&&\left\{a^3b^3\dot{\phi}\left[-\frac{1}{2}\beta_0+\beta_1\left(3H^2+3h^2+9Hh\right)
+\beta_2\left(36H^3h+36h^3H+108H^2h^2\right)-720\beta_3H^3h^3\right]\right\}_{,t}\nonumber\\&&+a^3b^3V_{,\phi}=0\;,\label{eom3}
\end{eqnarray}
where $H$ and $h$ are defined by
\begin{eqnarray}
H\equiv \frac{\dot{a}}{a}\;,\ \ \ \ h\equiv \frac{\dot{b}}{b}\;,
\end{eqnarray}
which could be understood as the Hubble parameters of universe A and universe B, respectively.
\section{A class of solutions}
\label{sec:4}
Observing the equations of motion (\ref{eom1},\ref{eom2},\ref{eom3}), we find there are four variables, $H,\ h,\ \phi,\ V$. But we have only three equations of motion. Thus the system of equations is not closed. One usually fix the expression of scalar potential in advance. However here, for simplicity, we shall assume the evolution of the scalar field as
\begin{eqnarray}
\phi=t\;,
\end{eqnarray}
which means the scalar field plays the role of cosmic time. By choosing different parameters, $\alpha_0,\ \alpha_1,\ \alpha_2,\ \alpha_3,\ \beta_0,\ \beta_1,\ \beta_2,\ \beta_3$ and different initial conditions, $H(0),\ h(0),\ V(0)$, we obtain a class of cosmic solutions as follows.

\textbf{A. Cyclic universes}

 When$\alpha_0=-3,\ \ \alpha_1=1,\ \ \beta_0=1$ (with other parameters vanishing) and $H(0)=1,\ \ h(0)=-1,\ \ V(0)=6$, we find both universe A and universe B undergo the cyclic evolution of expansion and contraction. Fig.~1 and Fig.~2 demonstrate the evolution of Hubble parameters and the scalar potential.

\textbf{B. Eternal expanding and contracting universe}

 When $\alpha_0=1,\ \ \alpha_1=1,\ \ \alpha_2=0.05,\ \ \beta_0=1$ (with other parameters vanishing) and $H(0)=1,\ \ h(0)=-1,\ \ V(0)=5$,   we find universe A is eternal expanding and universe B is eternal contracting. Fig.~3 and Fig.~4 demonstrate the evolution of Hubble parameters and the scalar potential.

\textbf{C. Two de Sitter phase}

  When $\alpha_0=1,\ \ \alpha_1=1,\ \ \alpha_2=0.05,\ \ \alpha_3=0.005,\ \ \beta_0=1,\ \ \beta_1=-0.1,\ \ \beta_2=0.22,\ \ \beta_3=-0.012$ and $H(0)=1,\ \ h(0)=-1,\ \ V(0)=5$, we find both universes evolve from one de Sitter phase to the other de Sitter phase. Fig.~5 and Fig.~6 demonstrate the evolution of Hubble parameters and the scalar potential. We see with the sharp increasing of scalar potential, transitions for the Hubble parameters are present.

\textbf{D. De Sitter solution}

\textbf{(a).} When $H=q,\ h=-q$ ($q>0$ is a constant), we obtain a constant scalar potential for the de Sitter solution
 \begin{eqnarray}
V=-\frac{1}{2}\alpha_0+3\alpha_1q^2-12\alpha_2q^4-144\alpha_3q^6-\frac{1}{4}\beta_0
-\frac{3}{2}\beta_1q^2+18\beta_2q^4+360\beta_3 q^6\;.
\end{eqnarray}
In this case, universe A is exponentially inflating while universe B is exponentially contracting.

\textbf{(b).} When $H=q,\ h=q$ ($q$ is a constant), we obtain the potential
 \begin{eqnarray}
V=-\frac{1}{2}\alpha_0+15\alpha_1q^2+180\alpha_2q^4-720\alpha_3q^6-\beta_0
+15\beta_1q^2+720\beta_3 q^6\;,
\end{eqnarray}
with
 \begin{eqnarray}
\beta_2=\frac{1440\beta_3q^6-30\beta_1q^2+\beta_0}{360q^4}\;.
\end{eqnarray}
In this case, both universes inflate exponentially.

\textbf{(c).} When $H=s,\ h=q$ ($s\neq q$), we obtain the potential
 \begin{eqnarray}
V&=&-24\beta_2 q^3 s-42\beta_2 s^2 q^2-2\beta_1q s-24s^3\beta_2 q-\frac{3}{2}\beta_1 q^2+3\alpha_1 s^2+
16\alpha_2 sq^3\nonumber\\&&+16\alpha_2s^3q+3\alpha_1q^2-\frac{1}{2}\alpha_0+4\alpha_1 sq+28\alpha_2s^2q^2-\frac{1}{4}\beta_0-\frac{3}{2}\beta_1s^2\;,
\end{eqnarray}
with
 \begin{eqnarray}
\alpha_3&=&\frac{1}{576s^3q^3}\left(16\alpha_2s^3q+16\beta_1qs+48s^3\beta_2q+120\beta_2s^2q^2+64\alpha_2s^2q^2+6\beta_1s^2\right.\nonumber\\&&\left.
+4\alpha_1sq+16\alpha_2sq^3+48\beta_2q^3s-\beta_0+6\beta_1q^2\right)\;,\\
\beta_3&=&-\frac{1}{1440s^3q^3}\left(-72s^3\beta_2q-72\beta_2q^3s-6\beta_1s^2-216\beta_2s^2q^2-18\beta_1qs+\beta_0-6\beta_1q^2\right)\;.
\end{eqnarray}
In this case, the two universes can inflate with arbitrary speed.

\textbf{E. Loop solution}

 When $\alpha_0=-1,\ \ \alpha_1=1,\ \ \alpha_2=0.005,\ \ \alpha_3=0.001,\ \ \beta_0=1,\ \ \beta_1=-0.01,\ \ \beta_2=0,\ \ \beta_3=0$ and $H(0)=1,\ \ h(0)=-1,\ \ V(0)=3.1$,  we find the trajectories of evolution for Hubble parameters forms a loop. This reveals universe A and universe B have the same Hubble scales at the beginning and the last. Fig.~7 and Fig.~8 demonstrate the evolution of Hubble parameters and the scalar potential. Also, the presence of loop is resulting from the bulging of the scalar potential.

\begin{figure}[h]
\begin{center}
\includegraphics[width=9cm]{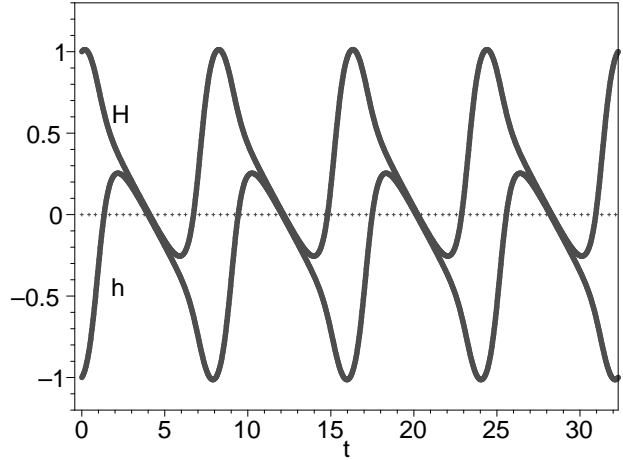}
\caption{Cyclic universes.}\label{fig:3}
\end{center}
\end{figure}

\begin{figure}[h]
\begin{center}
\includegraphics[width=9cm]{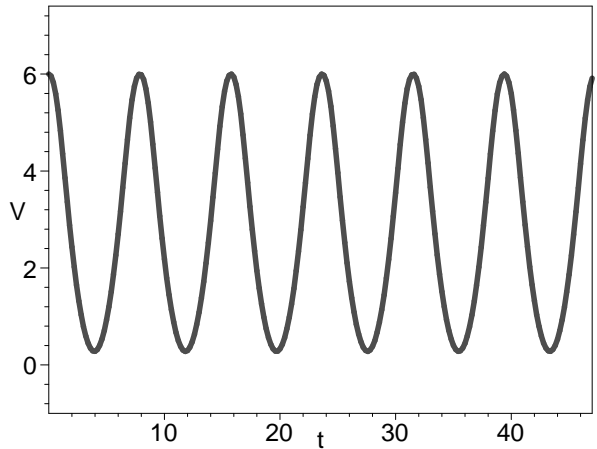}
\caption{The evolution of scalar potential in the cyclic-universes.}\label{fig:3}
\end{center}
\end{figure}

\begin{figure}[h]
\begin{center}
\includegraphics[width=9cm]{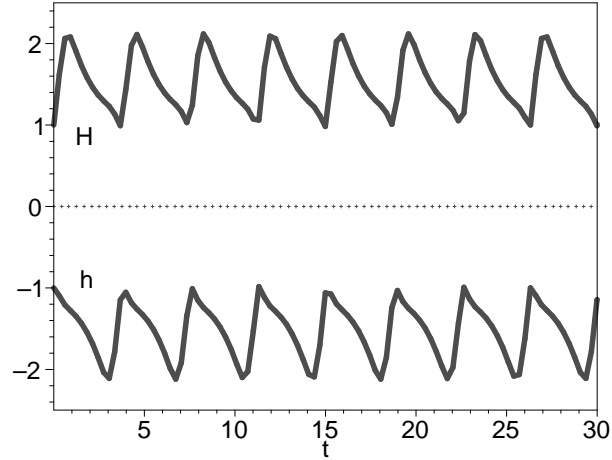}
\caption{Eternal expanding and contracting universes.}\label{fig:3}
\end{center}
\end{figure}

\begin{figure}[h]
\begin{center}
\includegraphics[width=9cm]{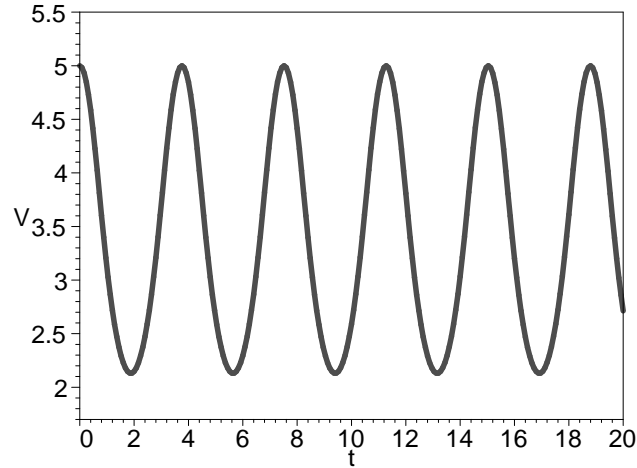}
\caption{The evolution of scalar potential in the eternal expanding and contracting universes.}\label{fig:3}
\end{center}
\end{figure}

\begin{figure}[h]
\begin{center}
\includegraphics[width=9cm]{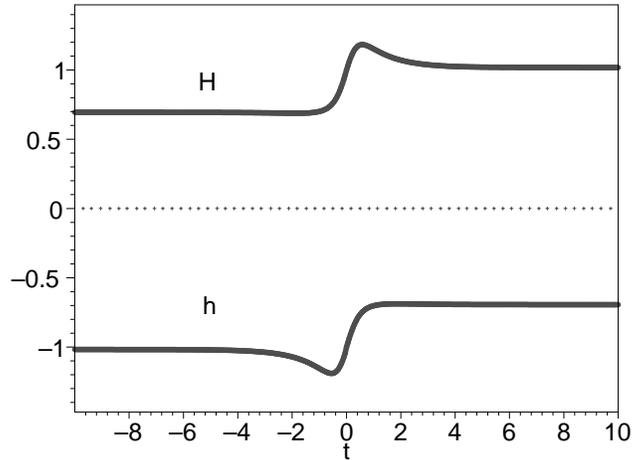}
\caption{The two-de Sitter-phase solution}\label{fig:3}
\end{center}
\end{figure}

\begin{figure}[h]
\begin{center}
\includegraphics[width=9cm]{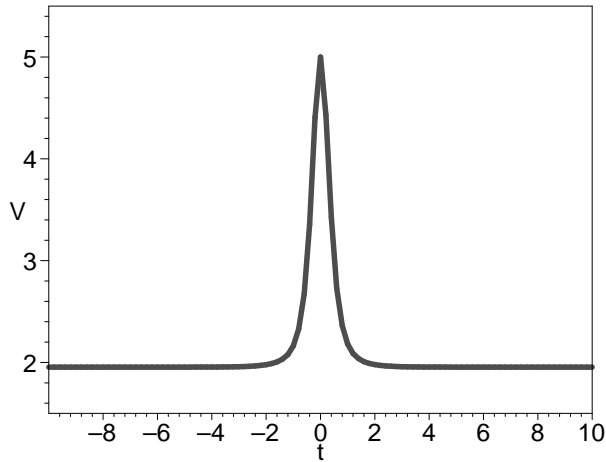}
\caption{The evolution of scalar potential for the two-de Sitter phase solution.}\label{fig:3}
\end{center}
\end{figure}

\begin{figure}[h]
\begin{center}
\includegraphics[width=9cm]{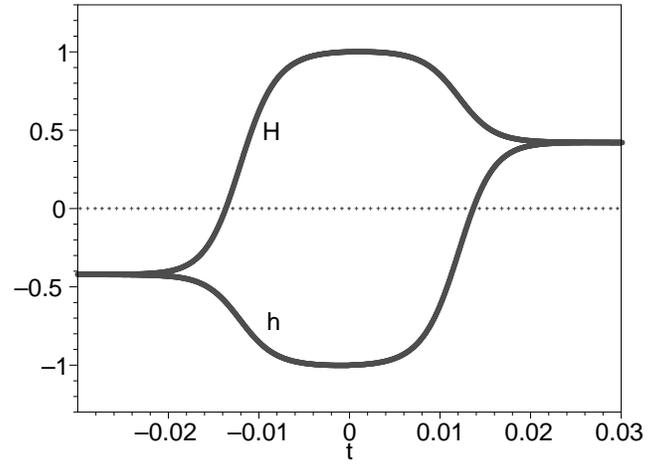}
\caption{The loop solution.}\label{fig:3}
\end{center}
\end{figure}

\begin{figure}[h]
\begin{center}
\includegraphics[width=9cm]{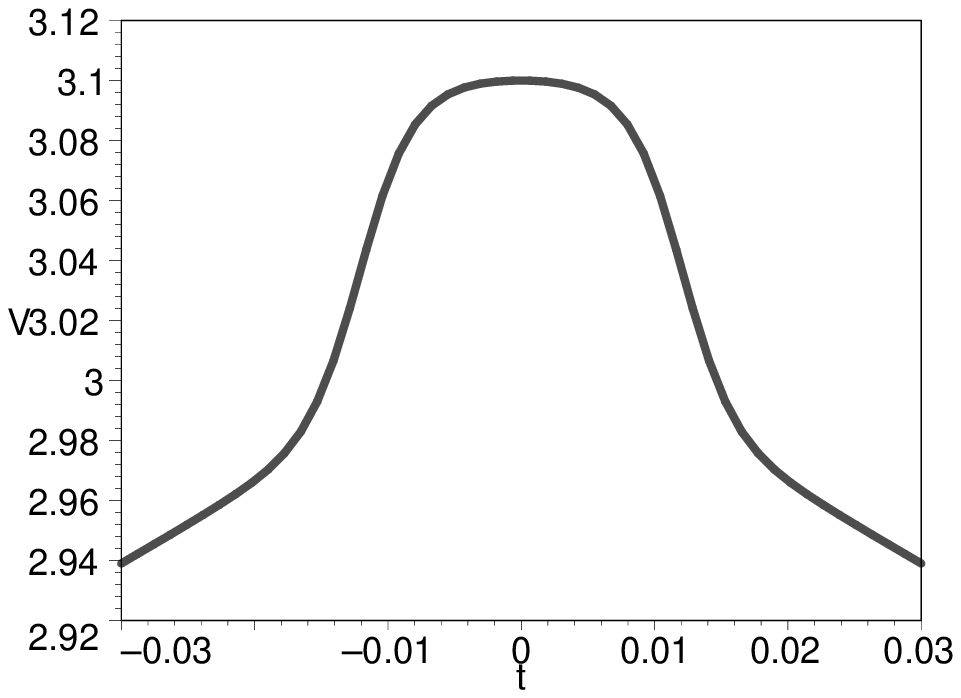}
\caption{The evolution of scalar potential for the loop solution.}\label{fig:3}
\end{center}
\end{figure}
\section{Discussion and conclusion}\label{sec:5}
In conclusion, we extend the Horndeski theories into the Lovelock gravity theory. The resulting equations of motion remain second order. As a subclass of the new Horndeski theories, the Lovelock scalar field is defined and applied into the investigation of cosmology. For simplicity, we pay attention to the $7$ dimensional spacetime where only four Lovelock tensors, $G_{\mu\nu}^{(0)},\ G_{\mu\nu}^{(1)},\ G_{\mu\nu}^{(2)},\ G_{\mu\nu}^{(3)}$  are in the presence. We find the physics of corresponding cosmology is rather rich such that there are a number of solutions. They are cyclic solution, eternal expanding or contacting solution, two-de Sitter solution, de Sitter solution and loop solution.

There are $9$ degree of freedom, $\alpha_i,\ \beta_i$ ($i=0,\ 1,\ 2,\ 3$) and $\phi(t)$, in the Lovelock scalar field model. We have assumed $\phi(t)=t$ in the cosmic evolution which means $\phi$ plays the role of cosmic time. Different assumptions on $\phi(t)$ would lead to different cosmic evolutions. Thus it deserves to be studied further. On the other hand, it is also important to look for the higher dimensional black hole solutions in this new Horndeski theories due to ADS/CFT.

\section*{Acknowledgments}
This work is partially supported by China Program of International ST Cooperation 2016YFE0100300
, the Strategic Priority Research Program ``Multi-wavelength Gravitational Wave Universe'' of the
CAS, Grant No. XDB23040100, the Joint Research Fund in Astronomy (U1631118), and the NSFC
under grants 11473044, 11633004, 11773031 and the Project of CAS, QYZDJ-SSW-SLH017.

\newcommand\ARNPS[3]{~Ann. Rev. Nucl. Part. Sci.{\bf ~#1}, #2~ (#3)}
\newcommand\AL[3]{~Astron. Lett.{\bf ~#1}, #2~ (#3)}
\newcommand\AP[3]{~Astropart. Phys.{\bf ~#1}, #2~ (#3)}
\newcommand\AJ[3]{~Astron. J.{\bf ~#1}, #2~(#3)}
\newcommand\APJ[3]{~Astrophys. J.{\bf ~#1}, #2~ (#3)}
\newcommand\APJL[3]{~Astrophys. J. Lett. {\bf ~#1}, L#2~(#3)}
\newcommand\APJS[3]{~Astrophys. J. Suppl. Ser.{\bf ~#1}, #2~(#3)}
\newcommand\JHEP[3]{~JHEP.{\bf ~#1}, #2~(#3)}
\newcommand\JMP[3]{~J. Math. Phys. {\bf ~#1}, #2~(#3)}
\newcommand\JCAP[3]{~JCAP {\bf ~#1}, #2~ (#3)}
\newcommand\LRR[3]{~Living Rev. Relativity. {\bf ~#1}, #2~ (#3)}
\newcommand\MNRAS[3]{~Mon. Not. R. Astron. Soc.{\bf ~#1}, #2~(#3)}
\newcommand\MNRASL[3]{~Mon. Not. R. Astron. Soc.{\bf ~#1}, L#2~(#3)}
\newcommand\NPB[3]{~Nucl. Phys. B{\bf ~#1}, #2~(#3)}
\newcommand\CMP[3]{~Comm. Math. Phys.{\bf ~#1}, #2~(#3)}
\newcommand\CQG[3]{~Class. Quantum Grav.{\bf ~#1}, #2~(#3)}
\newcommand\PLB[3]{~Phys. Lett. B{\bf ~#1}, #2~(#3)}
\newcommand\PRL[3]{~Phys. Rev. Lett.{\bf ~#1}, #2~(#3)}
\newcommand\PR[3]{~Phys. Rep.{\bf ~#1}, #2~(#3)}
\newcommand\PRd[3]{~Phys. Rev.{\bf ~#1}, #2~(#3)}
\newcommand\PRD[3]{~Phys. Rev. D{\bf ~#1}, #2~(#3)}
\newcommand\RMP[3]{~Rev. Mod. Phys.{\bf ~#1}, #2~(#3)}
\newcommand\SJNP[3]{~Sov. J. Nucl. Phys.{\bf ~#1}, #2~(#3)}
\newcommand\ZPC[3]{~Z. Phys. C{\bf ~#1}, #2~(#3)}
\newcommand\IJGMP[3]{~Int. J. Geom. Meth. Mod. Phys.{\bf ~#1}, #2~(#3)}
\newcommand\IJMPD[3]{~Int. J. Mod. Phys. D{\bf ~#1}, #2~(#3)}
\newcommand\GRG[3]{~Gen. Rel. Grav.{\bf ~#1}, #2~(#3)}
\newcommand\EPJC[3]{~Eur. Phys. J. C{\bf ~#1}, #2~(#3)}
\newcommand\PRSLA[3]{~Proc. Roy. Soc. Lond. A {\bf ~#1}, #2~(#3)}
\newcommand\AHEP[3]{~Adv. High Energy Phys.{\bf ~#1}, #2~(#3)}
\newcommand\Pramana[3]{~Pramana.{\bf ~#1}, #2~(#3)}
\newcommand\PTP[3]{~Prog. Theor. Phys{\bf ~#1}, #2~(#3)}
\newcommand\APPS[3]{~Acta Phys. Polon. Supp.{\bf ~#1}, #2~(#3)}
\newcommand\ANP[3]{~Annals Phys.{\bf ~#1}, #2~(#3)}

\end{document}